# CGB − A UNIX SHELL PROGRAM TO CREATE CUSTOM INSTANCES OF THE UCSC GENOME BROWSER

Vincenzo Forgetta[1*] & Ken Dewar[1]

[1]Department of Human Genetics, McGill University, Montreal, Quebec, Canada.

*Corresponding author: Lady Davis Institute for Medical Research, Jewish General Hospital, 3755 Côte Ste-Catherine Road, H-465 (Pavilion H), Montreal, Québec, H3T 1E2, Tel: +1 (514) 340-8222 ext.3982, Fax: +1 (514) 340-7564, Email: vincenzo.forgetta@mail.mcgill.ca

**Abstract**

The UCSC Genome Browser is a popular tool for the exploration and analysis of reference genomes. Mirrors of the UCSC Genome Browser and its contents exist at multiple geographic locations, and this mirror procedure has been modified to support genome sequences not maintained by UCSC and generated by individual researchers. While straightforward, this procedure is lengthy and tedious and would benefit from automation, especially when processing many genome sequences. We present a Unix shell program that facilitates the creation of custom instances of the UCSC Genome Browser for genome sequences not being maintained by UCSC. It automates many steps of the browser creation process, provides password protection for each browser instance, and automates the creation of basic annotation tracks. As an example we generate a custom UCSC Genome Browser for a bacterial genome obtained from a massively parallel sequencing platform.

**Introduction**

In the past, large institutions sequenced *de novo* the genome of organisms such as human (Lander et al., 2001), mouse (Waterston et al., 2002) and fly (Adams et al., 2000), and bioinformatics tools were created to provide the scientific research community with access to analyzing these reference genome resources. For instance, the scientific community routinely uses genome browsers to visualize and analyze a reference genome's sequence and annotations, with popular browsers being the UCSC Genome Browser (Kent et al., 2002), the NCBI Map Viewer (NCBI, 2011) and the Ensembl Genome Browser (Hubbard et al., 2002). These browsers provide a common set of functionality, such as data visualization and text searches, but each also offers unique functionalities. For example, the UCSC Genome Browser has tight integration with the BLAT sequence alignment tool (Kent, 2002), advanced database searching with the UCSC Table Browser (Karolchik et al., 2004), and extensibility via custom annotation tracks (http://genome.ucsc.edu/FAQ/FAQcustom.html). These features make this browser a leading resource for the analysis of over 40 reference genomes. As of mid-2012, the UCSC Genome Browser receives over 600,000 hits per day (http://genome.ucsc.edu/admin/stats/, accessed 28/05/12), and has been cited in more than 2,000 peer-reviewed articles.

Recently, massively parallel DNA sequencing (MPS) technologies (Bentley et al., 2008; Margulies et al., 2005; Shendure et al., 2005) have reduced the cost and time of genome sequencing dramatically, allowing individual researchers to sequence the entire genome of many organisms. However, the UCSC Genome Browser has remained primarily a community-based resource, thus excluding individual researchers from using this tool in their analysis of their own particular genome sequences. Recently, the mirror site installation procedure for the UCSC Genome Browser (http://genome.ucsc.edu/admin/mirror.html) has been modified to support non-reference genome sequences (http://genomewiki.ucsc.edu/index.php/Minimal_Browser_ Installation), but the procedure is lengthy and tedious, making it cumbersome to perform for many genome sequences.

We have created a Unix shell program, *cgb*, which facilitates the creation of custom instances of the UCSC Genome Browser. Each browser instance can be password-protected and can contain multiple genome sequences. We also include functionality to build genome sequences from contigs or scaffolds, and to automatically create basic annotations. Here we briefly describe the implementation and functionality of *cgb,* and provide an example usage case. The source code and documentation is available at http://github.com/vforget/cgb.

**Methods**

*Cgb* is written in the bash (Bourne-Again shell) scripting language. We chose this language because of its ubiquity on Unix-like platforms and its ability to execute the external programs required to setup a custom instance of the UCSC Genome Browser. *Cgb* relies on a functional installation of the UCSC Genome Browser (for more information see INSTALL.txt that is provided with *cgb*), but it does not require reference genome sequences or annotations. Each custom browser instance is secured using an Apache's distributed configuration file (.htaccess file).

We also provide programs to automate the building of a genome sequence from a set of contigs or scaffolds, and to create browser tracks for contigs, scaffolds, gaps, depth of read coverage, and GC content. This functionality was developed in the Python programming language.

**Results**

*General Functionality*

*Cgb* is a Unix shell program that presents the user with a series of tasks, with each task pertaining to a particular step in the browser installation process (Table 1). Prior to executing one or more tasks, the user specifies a new or existing identifier for the browser instance by setting the CLIENT_NAME variable (for more detail see Example Usage Case).

Table 1. List of *cgb* tasks and their commands.

| Task | Command(s) |
|---|---|
| Manage an instance of a UCSC Genome Browser | add, remove |
| Manage "Clade" entries | add, remove, list |
| Manage "Genome" entries for a particular "Clade" | add, remove, list |
| Manage "Build" entries for a particular "Genome" | add, remove, list |
| Add a FASTA a sequence for a Genome Build | add |
| Manage default Genome Builds | add, remove, list |
| Manage BLAT servers for a Genome Build | add, remove, list, restart |
| Add a contig assembly to a Genome Build | add |
| Add a scaffold assembly to a Genome Build | add |
| Add a depth of coverage annotation track to a Genome Build | add |

Each task has a set of commands (*add*, *remove*, *list*, or *restart*) to manage data entries that describe each genome sequence (i.e., clade, genome, and build) or that load a genome sequence and/or annotations into the browser instance (fasta, contig, scaffold, etc.). The *list* and *remove* commands are useful in cases where errors are committed or data is no longer required. Each command has a required set of arguments that specify the value of database entries (e.g., build name) or other properties pertaining to a genome sequence. A full description of these arguments is available via the *cgb* help message or in the documentation.

We have also provided extra commands that automate the converting of contigs or scaffolds into a genome build. By default, contigs or scaffolds are sorted by decreasing length and merged into one sequence record. *Cgb* also accepts contigs/scaffolds in a user-defined order (i.e. orthologous order on a related genome).

*Example Usage Case*

To create a custom instance of the UCSC Genome Browser we would execute the *cgb* commands listed in Figure 1.

The create_browser command adds a new instance (identified by ExampleClient1) of the UCSC Genome Browser, and protects the browser instance by prompting the user to set a username and password (not shown). The subsequent 4 steps (add_clade, add_genome, add_build, and

add_defaultdb) created entries in the browser's database. The add_fasta command loads a genome sequence in FASTA format, and add_blat starts the BLAT servers for this genome sequence.

Additional genome sequences can be added to the same browser instance (within or outside the same clade and/or genome), or a new browser instances can be created by setting a new CLIENT_NAME and repeating the tasks in Figure 1, but with different values.

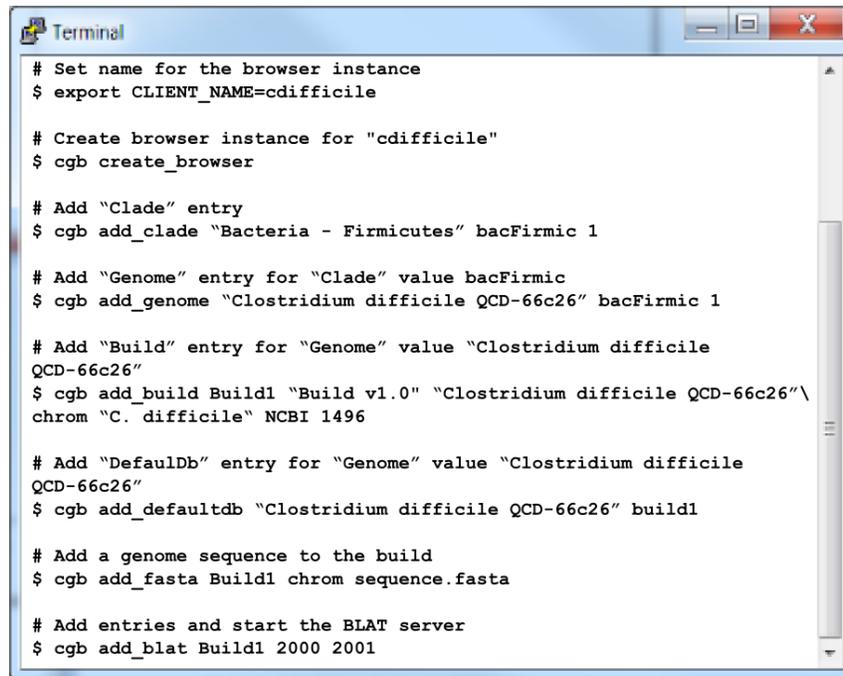

Figure 1. List of *cgb* commands for creating a custom UCSC Genome Browser.

Example describes the commands for creating a custom genome browser for the *C. difficile* isolate QCD-66c26 chromosome (CM000441.2). Lines beginning with a pound (#) or dollar sign ($) are comments or commands, respectively.

## Discussion

By converting a lengthy and tedious procedure into a series of simple tasks, *cgb* makes it easier to create custom instances of the UCSC Genome Browser. It has been used to support studies ranging from bacteria (Forgetta et al., 2011; Forgetta et al., 2012), to fungi (http://www.fungalgenomics.ca/wiki/Main_Page), to chordates (Matsumoto et al., 2010), and to mammals (Freimer et al., 2007). Because many of the tasks are non-interactive they can be further automated and customized by wrapping them into another Unix shell program, or more interestingly, through a web-browser interface, allowing users with no knowledge of Unix-like operating systems to create browser instances for non-reference genome sequences. *Cgb* demonstrates that existing bioinformatics tools can be adapted to address changes caused by MPS technologies, thereby reducing resources needed to develop a new tool.

## Availability and Requirements

This software is open source and is available for free at http://github.com/vforget/cgb. *Cgb* requires a pre-existing minimal install of the UCSC Genome Browser.

## Acknowledgements

I thank Pascale Marquis, Gary Leveque, and Jessica Wasserscheid for user testing and bug reporting. V.F. was the recipient of a Canadian Institutes of Health Research Doctoral Research Award. We are grateful to Jim Kent and the UCSC Genome Browser staff for developing and maintaining the UCSC Genome Browser source code and providing feedback during the development of *cgb*.

## References

Adams, M. D., Celniker, S. E., Holt, R. A., Evans, C. A., Gocayne, J. D., Amanatides, P. G., et al. (2000). The genome sequence of *Drosophila melanogaster*. *Science, 287*(5461), 2185-2195.

Bentley, D. R., Balasubramanian, S., Swerdlow, H. P., Smith, G. P., Milton, J., Brown, C. G., et al. (2008). Accurate whole human genome sequencing using reversible terminator chemistry. *Nature, 456*(7218), 53-59.

Forgetta, V., Oughton, M. T., Marquis, P., Brukner, I., Blanchette, R., Haub, K., et al. (2011). Fourteen-genome comparison identifies DNA markers for severe-disease-associated strains of *Clostridium difficile*. *Journal of clinical microbiology, 49*(6), 2230-2238.

Forgetta, V., Rempel, H., Malouin, F., Vaillancourt, R., Jr., Topp, E., Dewar, K., et al. (2012). Pathogenic and multidrug-resistant *Escherichia fergusonii* from broiler chicken. *Poultry science, 91*(2), 512-525.

Freimer, N. B., Service, S. K., Ophoff, R. A., Jasinska, A. J., McKee, K., Villeneuve, A., et al. (2007). A quantitative trait locus for variation in dopamine metabolism mapped in a primate model using reference sequences from related species. *Proc Natl Acad Sci U S A, 104*(40), 15811-15816.

Hubbard, T., Barker, D., Birney, E., Cameron, G., Chen, Y., Clark, L., et al. (2002). The Ensembl genome database project. *Nucleic acids research, 30*(1), 38-41.

Karolchik, D., Hinrichs, A. S., Furey, T. S., Roskin, K. M., Sugnet, C. W., Haussler, D., et al. (2004). The UCSC Table Browser data retrieval tool. *Nucleic Acids Res, 32*(Database issue), D493-496.

Kent, W. J. (2002). BLAT--the BLAST-like alignment tool. *Genome Res, 12*(4), 656-664.

Kent, W. J., Sugnet, C. W., Furey, T. S., Roskin, K. M., Pringle, T. H., Zahler, A. M., et al. (2002). The human genome browser at UCSC. *Genome Res, 12*(6), 996-1006.

Lander, E. S., Linton, L. M., Birren, B., Nusbaum, C., Zody, M. C., Baldwin, J., et al. (2001). Initial sequencing and analysis of the human genome. *Nature, 409*(6822), 860-921.

Margulies, M., Egholm, M., Altman, W. E., Attiya, S., Bader, J. S., Bemben, L. A., et al. (2005). Genome sequencing in microfabricated high-density picolitre reactors. *Nature, 437*(7057), 376-380.

Matsumoto, J., Dewar, K., Wasserscheid, J., Wiley, G. B., Macmil, S. L., Roe, B. A., et al. (2010). High-throughput sequence analysis of Ciona intestinalis SL trans-spliced mRNAs: alternative expression modes and gene function correlates. *Genome Res, 20*(5), 636-645.

NCBI. (2011). NCBI Map Viewer  Retrieved December 20, 2011, 2011, from http://www.ncbi.nlm.nih.gov/projects/mapview/

Shendure, J., Porreca, G. J., Reppas, N. B., Lin, X., McCutcheon, J. P., Rosenbaum, A. M., et al. (2005). Accurate multiplex polony sequencing of an evolved bacterial genome. *Science, 309*(5741), 1728-1732.

Waterston, R. H., Lindblad-Toh, K., Birney, E., Rogers, J., Abril, J. F., Agarwal, P., et al. (2002). Initial sequencing and comparative analysis of the mouse genome. *Nature, 420*(6915), 520-562.